\documentclass[conference]{IEEEtran}
\IEEEoverridecommandlockouts
\usepackage[noadjust]{cite}

\usepackage{amsmath,amssymb,amsfonts}
\usepackage{algorithmic}
\usepackage{graphicx}
\usepackage{textcomp}
\usepackage{xcolor}
\usepackage{hyperref}
\usepackage{enumitem}
\def\BibTeX{{\rm B\kern-.05em{\sc i\kern-.025em b}\kern-.08em
    T\kern-.1667em\lower.7ex\hbox{E}\kern-.125emX}}

\makeatletter
 \let\old@ps@headings\ps@headings
 \let\old@ps@IEEEtitlepagestyle\ps@IEEEtitlepagestyle
 
 \def\confheader#1{%
  \def\ps@IEEEtitlepagestyle{%
    \old@ps@IEEEtitlepagestyle%
    \def\@oddhead{\strut\hfill#1\hfill\strut}%
    \def\@evenhead{}%
  }%
  \def\ps@headings{%
    \old@ps@headings%
    \def\@oddhead{}%
    \def\@evenhead{}%
  }%
 \ps@headings%
 }

 \makeatother

\confheader{
Accepted in 2025 IEEE 15th Annual Computing and Communication Workshop and Conference (CCWC)
}

\usepackage[pscoord]{eso-pic}
\newcommand{\placetextbox}[3]{
 \setbox0=\hbox{#3}
 \AddToShipoutPictureFG*{ \put(\LenToUnit{#1\paperwidth},\LenToUnit{#2\paperheight}){\vtop{{\null}\makebox[0pt][c]{#3}}}
 }
 }
 \placetextbox{.21}{0.055}{\small{979-8-3315-0769-5/25/\$31.00~\copyright 2025 IEEE}}

\begin{document}

\title{Leveraging Large Language Models and Machine Learning for Smart Contract Vulnerability Detection}
\author{
    \IEEEauthorblockN{S M Mostaq Hossain, Amani Altarawneh and Jesse Roberts}
    \IEEEauthorblockA{\textit{Department of Computer Science}, \textit{Tennessee Technological University}\\
    Cookeville, Tennessee, USA \\
    Email: \{shossain42, aaltarawneh, 
jtroberts\}@tntech.edu}
}

\maketitle

\begin{abstract}
As blockchain technology and smart contracts become widely adopted, securing them throughout every stage of the transaction process is essential. The concern of improved security for smart contracts is to find and detect vulnerabilities using classical Machine Learning (ML) models and fine-tuned Large Language Models (LLM). The robustness of such work rests on a labeled smart contract dataset that includes annotated vulnerabilities on which several LLMs alongside various traditional machine learning algorithms such as DistilBERT model is trained and tested. We train and test machine learning algorithms to classify smart contract codes according to vulnerability types in order to compare model performance. Having fine-tuned the LLMs specifically for smart contract code classification should help in getting better results when detecting several types of well-known vulnerabilities, such as Reentrancy, Integer Overflow, Timestamp Dependency and Dangerous Delegatecall. From our initial experimental results, it can be seen that our fine-tuned LLM surpasses the accuracy of any other model by achieving an accuracy of over 90\%, and this advances the existing vulnerability detection benchmarks. Such performance provides a great deal of evidence for LLMs' ability to describe the subtle patterns in the code that traditional ML models could miss. Thus, we compared each of the ML and LLM models to give a good overview of each model's strengths, from which we can choose the most effective one for real-world applications in smart contract security. Our research combines machine learning and large language models to provide a rich and interpretable framework for detecting different smart contract vulnerabilities, which lays a foundation for a more secure blockchain ecosystem.

\end{abstract}


\section{Introduction}
The increasingly rapid evolution of blockchain technology has fostered the large-scale adoption of smart contracts, defined as self-executing contracts with the terms of the accord explicitly written into code. With various application areas of smart contracts such as finance, supply chain management, and decentralized applications (DApps) rising quite spectacularly, their securing has become a very critical aspect \cite{Kiani_Sheng_2024}. Security vulnerabilities of smart contracts can lead to dire consequences, including financial losses, breaches of privacy, and systematic risks posed to decentralized systems \cite{Surucu_2022}. The growing prevalence of incidents within the blockchain arena shows the critical consequences of these vulnerabilities, making it clear that effective means of detection for such vulnerabilities need to exist.

Detection of vulnerabilities within smart contracts has mostly relied on static analysis techniques \cite{ghaleb2020effective} that analyze the code of the contract without executing it. These techniques can detect known vulnerabilities but have faced difficulties in handling rising complications alongside emerging vectors of attack. Thus, alternative approaches, such as machine learning and deep learning, are being explored by researchers to assist in the detection of various types of attacks \cite{Sun_Gu_2021}). These machine learning and deep learning-based methods use training algorithms that can learn from annotated datasets to find patterns that could reveal security holes. This makes it easier to find both known and unknown security problems \cite{zhou2022vulnerability, tang2023deep}.

The integration of large language models in the context of vulnerability detection for smart contracts signifies a notable advancement in the field of blockchain cybersecurity \cite{boi2024smart}. Models that command the ability to generate and comprehend the code-in a contextual manner-seem qualified for outperforming the traditional ML models by capturing complicated relationships and semantic information within the contract code. Recent research \cite{kim2024robust} into the use of fine-tuned Transformer encoder models for effective vulnerability detection finds focus being directed ever more closely toward imposing language models on the enhancement of smart contract security. Building on these advancements, the current work explores how ML and LLM can enhance smart contract vulnerability detection. The main goal of this research is as follows: 




\begin{enumerate}[label=(\alph*)]
    \item Comparative Evaluation: A systematic comparison is carried out, contrasting classical ML models such as LSTM \cite{6795963} and fine-tuned large language models \cite{kim2024robust}, like DistilBERT \cite{sanh2019distilbert} and BERT \cite{devlin2018bert}, for prioritizing smart contract vulnerability detection and providing paramount effectiveness of LLMs in terms of capturing complex, context-rich patterns.
    \item Detection of Subtle Code Patterns: This study sets forth the capability of LLMs in determining intricate code patterns and dependencies-this classical model misses in vast applications, more prominently with regards to vulnerabilities like reentrancy \cite{samreen2020reentrancy} and integer overflow \cite{lai2020static} to show improved pattern recognition.
    \item Understanding the Impact of Fine-Tuning: We evaluate how fine-tuning LLMs on domain-specific smart contract data improves model performance and generalization, giving better precision and recall with respect to complex vulnerabilities.
    \item Benchmarking and Future Directions: In this work, we establish thorough benchmarks in vulnerability detection, expanding the scope of future research with an eye toward additional dataset expansion, wider LLM explorations, and the use of semi-supervised learning methods for improvement in generalization.
\end{enumerate}

The remaining paper is set out organized as follows: The next section explains the Key Research Questions followed by the Background that depicts the relevant work. The Methodology explains how our approach was executed, including data preparation and model evaluation. Results provide a comparison of classical ML models in the form of LLMs, followed by a Discussion of these findings. Future Work is suggested, and then we conclude with key takeaways in the Conclusions. 

\section{Key Research Questions}


The Ethereum blockchain ecosystem \cite{wohrer2018smart} relies on smart contracts security, the purpose of this research is to fix critical vulnerabilities in these contracts. To clarify and provide a framework for evaluation, following research questions lead the investigation. The questions test how well ML models and LLMs compare, whether an LLM fine-tuned to domain-specific data would benefit and whether LLMs could leverage minute code patterns that standard models wouldn't recognize.
\begin{itemize}
    \item \textbf{RQ-1:} How do classical machine learning models compare to fine-tuned large language models (LLMs) in smart contract vulnerability detection in terms of effectiveness?
    \item \textbf{RQ-2:} Are subtle code patterns and dependencies in smart contracts captured by LLMs, given that those very patterns and dependencies evade classical machine learning models?
    \item \textbf{RQ-3:} What are the effects of the fine-tuned large language model training on labeled smart contract data in enhancing vulnerability detection accuracy, and how do they affect the model’s generalization?
\end{itemize}

\section{Background}

Research on smart contract security has been exponentially growing over time. This concern is justified given the potential safety and security risks associated with blockchain technology. The earliest works within this domain focused on static analysis tools \cite{feist2019slither} for identifying known vulnerabilities without running the associated code. Researchers were able to leverage the static nature of smart contracts to analytically check for breaches and subsequently formalize them using mathematical proof systems. It is currently the approach followed in most vulnerability detection practices. This however requires manual routine inspection, like developers acquiring an understanding of contract logic in order to conduct the audit. Since the approach is "static", there is the natural tendency to obtain more false positives \cite{Yang_Zhang_Gu_Cui_2022} as opposed to dynamic results which is expected to be more precise and hence are assumed to be "dynamic". How do we trust, accept or understand security outcome that we are aware of it generates more false alarms?. This may have a resultant effect in the user acceptance of the security measure.

In order to address these shortcomings enumerated in the static approach to smart contract security checks. Numerous authors have since attempted to build ML models such as decision trees and support vector machines \cite{electronics13122295} to automatically detect vulnerabilities. These new algorithms take source code as input, extract features from it, build models using the extracted features and then use the model to ultimately classify whether or not a contract has vulnerabilities. Recent research \cite{Sun_Gu_2021} have demonstrated the efficacy of several models in identifying vulnerabilities such as reentrancy and gas limit concerns, attaining notable accuracy rates \cite{samreen2020reentrancy}. Nonetheless, numerous techniques continue to depend significantly on manually built features, constraining their scalability and responsiveness to emerging risks.

Recent research \cite{Prifti_Çiço_Karras_2024}, particularly employing Deep Learning for vulnerability detection. Researchers have also found that CNN and LSTM models can extract higher-level representations of smart contracts, leading to improved accuracy detection rates \cite{Wang_Zheng_Sun_2022}. Deep learning improves code analysis efficiency by automatically learning features from raw inputs, overcoming limitations in classical ML capabilities, particularly in feature engineering. Smart contracts are vulnerable to different attacks that compromise security and reliability \cite{he2023detection}. Common issues include Reentrancy (RE) \cite{samreen2020reentrancy}, Integer Overflow (IO) \cite{lai2020static}, Timestamp Dependence (TD) \cite{zhuang2021smart}, and Dangerous Delegatecall (DD) \cite{su2022effectively}, which can lead to arbitrary code execution.
Understanding these weaknesses is crucial for developing efficient detection algorithms.

Encoder-only models like BERT excel at language understanding because they have unfettered access to the full input context, making them ideal candidates for context-aware vulnerability detection in smart contract codes \cite{Gong_Song_Wang_Wang_Zhu_2023}. Researchers who adapt these algorithms to smart contract datasets have observed considerable improvements in vulnerability detection. In the paper \cite{ma2024combining} Ma et al. introduce an LLM prompt-based framework, iAudit, which aims to fine-tune LLMs for 263 real smart contract auditing through a two-stage decision-making and justification process. The model adds an iterative approach using LLM-based agents to refine the vulnerability explanations. An F1 score of 91.21\% is demonstrated together with superiority in detection and consistency relative to existing fine-tuned models such as CodeBERT \cite{feng2020codebert}.
Another paper \cite{boi2024smart} by Boi et al. utilizes the Llama-2-7b model for vulnerability detection and got an accuracy of 59.5\% with their Smartbugs Dataset (143 smart contracts). However, their approach demonstrates the potential of using LLMs for vulnerability detection and highlights an opportunity to further improve performance by expanding the dataset and employing enhanced fine-tuning strategies.



\begin{figure}
  \centering
  \includegraphics[width=\linewidth, keepaspectratio]{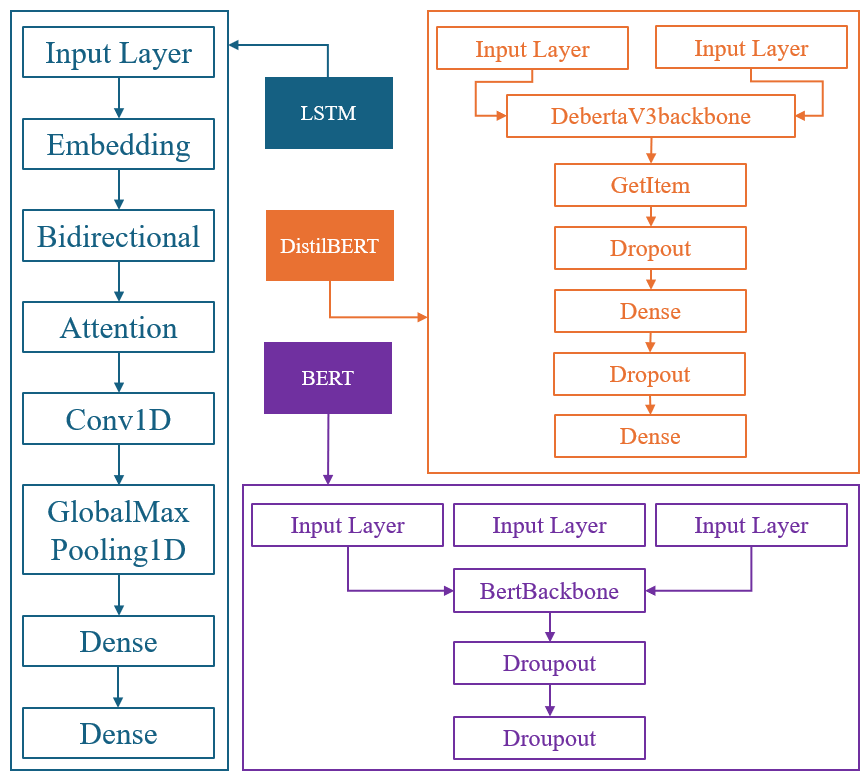}
  \caption{Customized layer structure of data flow diagram for LSTM, DistilBERT and BERT Models in our experiment.}
  \label{fig:model-str}
\end{figure}

\begin{figure*}
  \centering
  \includegraphics[width=\linewidth, keepaspectratio]{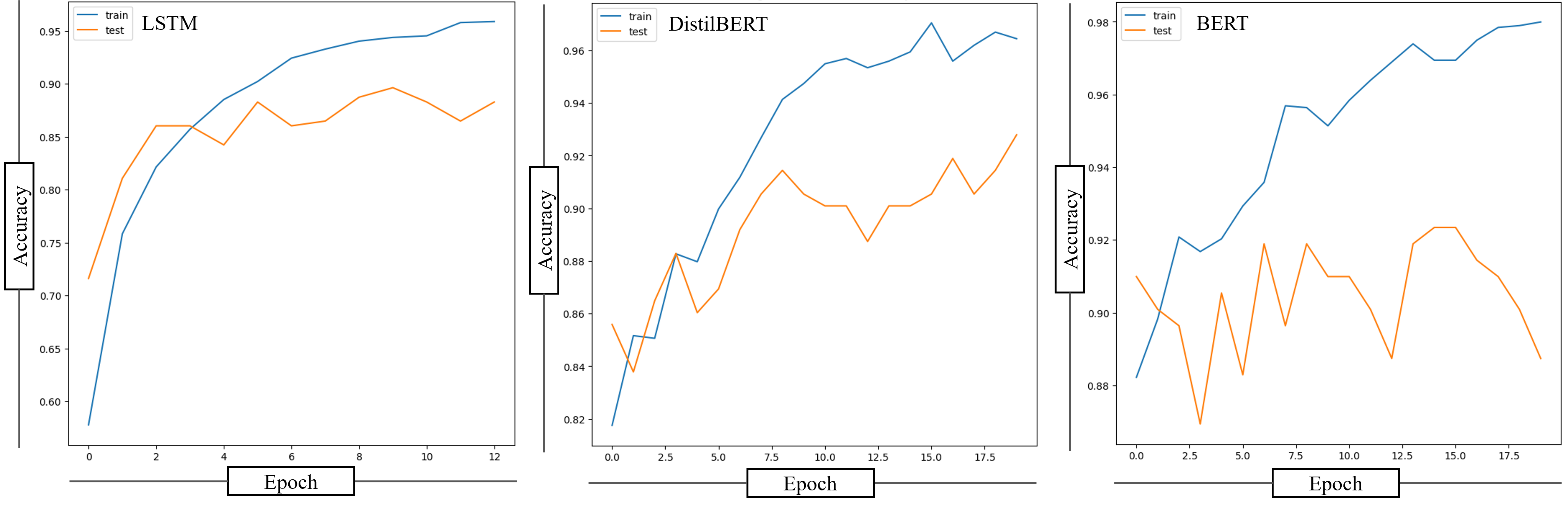}
  \caption{Training and validation accuracy plots for LSTM, DistilBERT, and BERT models over 20 epochs. All models demonstrate increasing accuracy during training, with BERT achieving the highest peak validation accuracy. However, BERT also exhibits the highest variance in validation accuracy.}
  \label{fig:acc-all}
\end{figure*}

\begin{figure*}
  \centering
  \includegraphics[width=\linewidth, keepaspectratio]{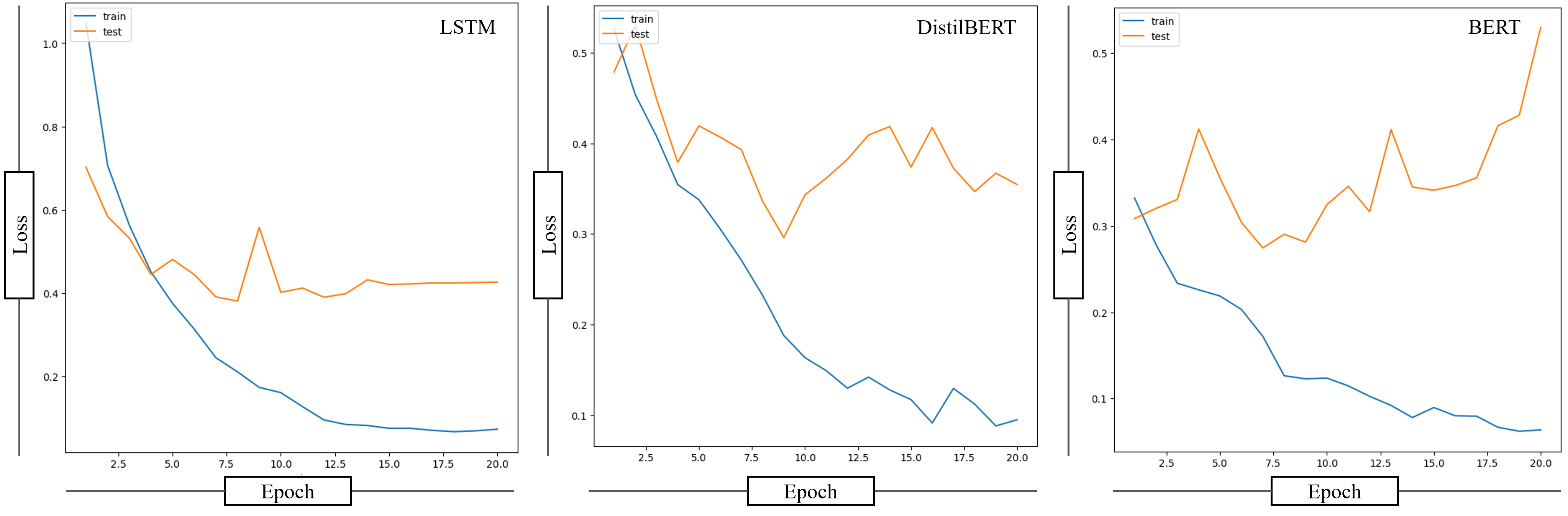}
  \caption{Training and validation loss plots for LSTM, DistilBERT, and BERT models over 20 epochs. While all models show a decreasing trend in validation loss, BERT consistently outperforms the other models in terms of minimum validation loss. However, BERT's validation loss also fluctuates more significantly, indicating potential instability in its learning process.}
  \label{fig:loss-all}
\end{figure*}

\begin{figure*}
  \centering
  \includegraphics[width=\linewidth, keepaspectratio]{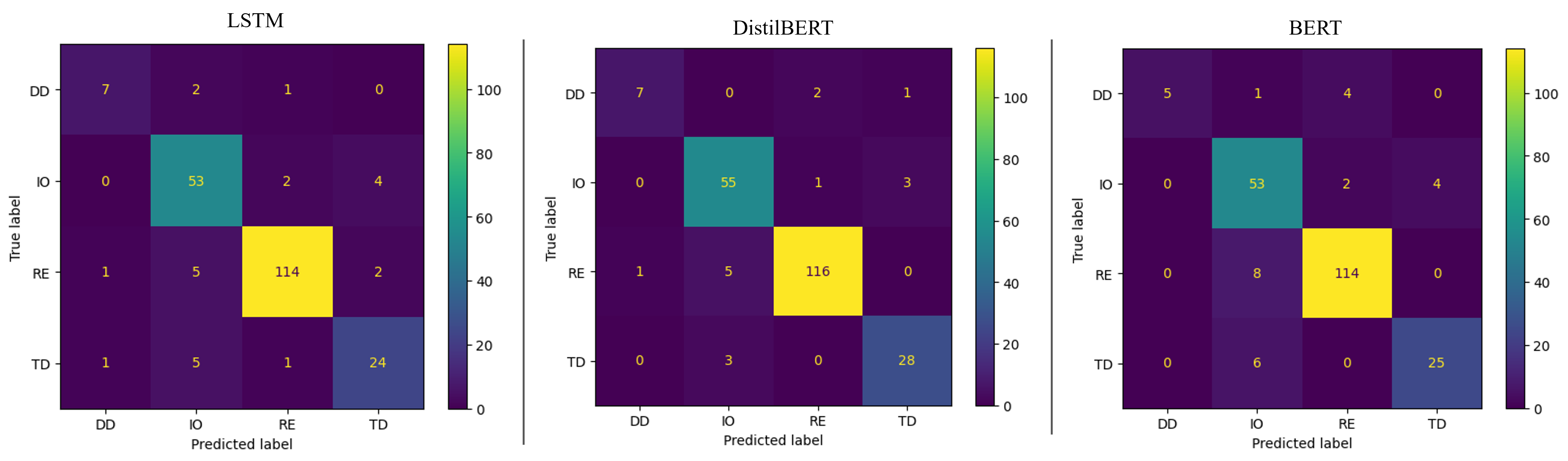}
  \caption{Confusion matrices for LSTM, DistilBERT, and BERT models, visualizing the distribution of true and predicted labels. All models demonstrate a strong bias toward predicting the Timestamp Dependency (TD) class, with BERT showing the highest accuracy for the Reentrancy (RE) class.}
  \label{fig:confmat-all}
\end{figure*}

\section{Methodology}
This section describes the procedure in an organized way that was adapted to combat the problem of smart contract vulnerability detection using ML and LLMs. It incorporates the dataset employed, models applied and an experimental setup for carrying out the analysis.
\subsection{Dataset Overview}\label{AA}
The primary dataset for this work consisted of an annotated smart contract dataset \cite{liu2023rethinking} from Liu et al. This dataset contains smart contracts annotated with various vulnerabilities, providing in-depth analysis and testing. It consists of 2,217 smart contracts with a structured representation of 5 columns: filename, code snippet, vulnerability label, and an encoded label. The dataset includes four major types of smart contract vulnerabilities: Reentrancy, which has the largest collection with 1,218 samples; Integer Overflow with 590 samples; Timestamp Dependency with 312 samples; and Dangerous Delegate Call with 97 samples. Each contract features the Solidity \cite{ferreira2020smartbugs} source code along with tagged annotation indicating the type of the specific vulnerability it presents, hence qualifying it for training and evaluation of machine learning and LLM models during vulnerability detection tasks.


\begin{table*}
\centering
\caption{Precision, recall, F1-score, and support for each vulnerability class (DD, IO, RE, TD) and overall macro and weighted averages for LSTM, DistilBERT, and BERT models.}
\resizebox{\textwidth}{!}{ 
\begin{tabular}{lcccccccc}
\hline
\\[-0.8em]
\textbf{Model} & \textbf{DD (Dangerous Delegatecall)} & \textbf{IO (Integer Overflow)} & \textbf{RE (Reentrancy)} & \textbf{TD (Timestamp Dependency)} \\
                & \textbf{Precision / Recall / F1 Score / Support} & \textbf{Precision / Recall / F1 Score / Support} & \textbf{Precision / Recall / F1 Score / Support} & \textbf{Precision / Recall / F1 Score / Support} \\
\\[-0.8em] \hline
\\[-0.8em]
LSTM           & 0.78 / 0.70 / 0.74 / 10  & 0.82 / 0.90 / 0.85 / 59  & 0.97 / 0.93 / 0.95 / 122 & 0.80 / 0.77 / 0.79 / 31 \\
DistilBERT     & 0.88 / 0.70 / 0.78 / 10  & 0.87 / 0.93 / 0.90 / 59  & 0.97 / 0.95 / 0.96 / 122 & 0.88 / 0.90 / 0.89 / 31 \\
BERT           & 1.00 / 0.50 / 0.67 / 10  & 0.78 / 0.90 / 0.83 / 59  & 0.95 / 0.93 / 0.94 / 122 & 0.86 / 0.81 / 0.83 / 31 \\
Accuracy       & 0.89                    & 0.93                    & 0.89                    & 0.93                   \\
Macro Avg      & 0.84 / 0.83 / 0.83 / 222 & 0.90 / 0.87 / 0.88 / 222 & 0.90 / 0.78 / 0.82 / 222 & 0.89 / 0.89 / 0.89 / 222 \\
Weighted Avg   & 0.89 / 0.89 / 0.89 / 222 & 0.93 / 0.93 / 0.93 / 222 & 0.89 / 0.89 / 0.89 / 222 & 0.89 / 0.89 / 0.89 / 222 \\
\\[-0.8em] \hline
\end{tabular}
}
\label{table:evaluation-metrics}
\end{table*}

\subsection{Models}
The first experiment used an LSTM-multi-layer RNN \cite{sherstinsky2020fundamentals}, which is good at processing data sequences. 
BERT was trained on a large corpus of textual data, including structured programs where DistilBERT is a smaller model distilled from BERT \cite{devlin2018bert}. Figure \ref{fig:model-str} shows the layered dataflow structure of the models which has been used during our experiment.


\begin{itemize}
    \item LSTM: The Long short-term memory (LSTMs) analyze linguistic input via recursion, adding each tokenized input to a latent representation that is ultimately used to represent information about the total context. RNN-based techniques are typically outperformed by encoder-only transformer-based techniques in many natural language understanding tasks due to the latter's ability to selectively attend to the entire context without the need for recursion. This makes them more able to capture long-range dependencies \cite{devlin2018bert}.  Therefore, we additionally test both BERT and the more efficient DistilBERT in the same application. However, for the LSTM model layer design Bidirectional LSTM with attention and convolution layers have been used \cite{LIU2019325}.
    \item DistilBERT: A distilled variant of the BERT \cite{sanh2019distilbert} language model is small, fast, cheap, and, hence, efficient, keeping over 90\% of BERT's language understanding efficacy. Efficiency earns it a great deal of applicability in the world of practical issues where computational resources are at a premium.
    \item BERT: In the experiment, Bidirectional Encoder Representations from Transformers (BERT) \cite{devlin2018bert} model has been used which changed the landscape of natural language processing. The architecture of BERT allows it to learn the deep contextual relationships in the data, making it a natural match for detecting well-known vulnerabilities from the smart contract code.
\end{itemize}

In this regard, the effectiveness of all models was thoughtfully assessed, followed by experience-based discussions aimed at charting the course to determine the approach to take care of the better smart contract vulnerability detection.

\subsection{Experimental Setup}

Experiments were run on a Google Colab notebook, which is an online Collaborative programming environment best suited for performing Computation heavy tasks. Using Google Colab has the additional benefit of access to GPUs, which speeds up training time for ML and LLM models. The steps of the experimental setup were:
\begin{itemize}
    \item Data Preparation: It was used to pre-process the data, clean it from code fragments and turn them into a way they can be fed into the models. These comprised of tokenization, normalizing the coding syntax and encoding vulnerabilities to a structured format.
    \item Hardware and Computational Resources: The experiments utilized Google Colab Pro with an NVIDIA A100 GPU (40 GB RAM), 83.5 GB system RAM, and 58.69 compute units at 8.47 units/hour, ensuring efficient model training and testing for vulnerability detection.
    \item Model Implementation: The implementation consisted of coding the models in python with appropriate libraries such as TensorFlow and Keras NLP models \cite{keras_nlp_models} to go through standard pre-training. Next, the models were trained on this dataset with respective hyperparameter tuning for optimal performance.
    \item Evaluation metrics: For this work, evaluation metrics such as accuracy, precision, recall, and the F1 score, allowed a quantitative assessment of the performance of the models in performing vulnerability detection. For comparison purposes, the results were documented.
    \item Iterative testing: It is provided by iterative assessments. As the testing progressed, feedback from the model performance helped adjust the dataset and model parameters, enabling continuous improvement and fine-tuning of detection.
\end{itemize}

With a proper methodology, it is natural to move forward toward the detection of smart contracts' vulnerabilities with the infusion of mixed traditions of machine-learning methods and large language models.

\section{Results}
The results are arranged in several critical findings stemming from the analyses performed by ML and LLMs, which exposed vulnerabilities from other smart contract syntactic-literal-based techniques. The evaluation compares the LSTM, DistilBERT, and BERT models on a labeled dataset of smart contracts, with annotated vulnerabilities shown in Table \ref{table:evaluation-metrics}. DistilBERT generally outperforms LSTM and BERT across almost all metrics, especially in terms of macro and weighted averages. This would imply that DistilBERT offers a more balanced performance across the different vulnerability classes. BERT achieved relatively high precision for the "DD" class but low recall and F1-score, indicating that it may be better at identifying actual true positives but constantly struggles against false positives. 

The LSTM model was classified among high-poised ones, giving it an overall accuracy of 89\% that ensured it was upheld by a reasonably rich classification report with sufficiently high precision (0.97 on the 'RE' class) and good recall on most classes. It indicates from the confusion matrix of Figure \ref{fig:confmat-all} that it performed particularly well by predicting the 'RE' class correctly (114 correct of 122) with only minor misclassifications in the other categories, such as 'IO' and 'TD.' In parallel, while the training loss noticeably decreased with every epoch, the validation loss became even more steady after the tenth epoch, assuring quite good generalization with very little overfitting.

The overall performance of the DistilBERT model was the highest, standing at 87\% (test loss: 0.296 in Figure \ref{fig:loss-all}), thus signifying good performance. From the provided classification report, the model did indeed score high marks in predicting the 'RE' class, attaining a precision of 0.98 and an F1 score of 0.96 while the performance in 'DC' and 'TD' classes was a lot less. Training progress indicates an upward trend with steady increases in accuracy through epochs, terminating at maximum training accuracy of 93.4\% although validating accuracy stabilized around 86-87\%. in which learning is quite effective although with some fluctuations in validation loss. 

The BERT model performed satisfactorily, achieving 89\% accuracy and 0.53-loss on their test in Figure \ref{fig:acc-all} and Figure \ref{fig:loss-all} . The maximum validation accuracy during training was recorded at 92.3\%. The classification report indicates particularly high F1-scores concerning the RE class, while the DD class performed far lesser at an F1-score of 0.67, likely due to underrepresented support. On the whole, the loss of the model showed the way down through the training period, except for the last few epochs during which the validation loss showed a steep rise, indicating a probable case of overfitting.

To elaborate further, the confusion matrix (BERT) has some class confusion, especially with class RE and class TD misclassifications, but overall, class RE is well classified. Another matrix (DeBERTa) has achieved improved accuracy across most categories, most notably with reduced errors in the DD and TD classes, whereas the LSTM is almost equal to BERT on prediction accuracies but has slightly more misclassifications in TD, indicating that it struggles in distinguishing that class. 

\section{Discussion}
This paper contrasts the issue of smart contract vulnerability detection, creating a comparative analysis through: Classical ML models (LSTM) and well-tuned LLMs (DistilBERT and BERT). Multiple valuable recommendations surfaced in tandem with research questions.

\begin{enumerate}[label=(\alph*)]
    \item The efficiency of Models (RQ-1): Obtaining an accuracy of 89\% with LSTM shows performable results that rival the other, like BERT, thereby offering a good performance. This same model shows regular training with low overfitting and  great generalizability in classifying ''RE" samples properly. Vulnerabilities that are more context dependent like "TD" were better covered by LLMs. DistilBERT provided relatively better results in demanding ''RE,'' with the maximum accuracy of 87\% and F1-scores. BERT achieved 89\% accuracy, but with overfitting problems losing validation at later epochs. The LLMs pattern more complex than LSTM in complicated pattern capturing and better able to accommodate delicate context-rich code dependencies.
    \item Code Patterns Capture (RQ-2): LLMs were thus shown under to be superior at detecting hidden patterns within the smart contracts, utilizing their already-trained knowledge to detect loopholes one thing which the LSTM may fail to do. This showed itself with better precision on the 'RE' class, having a greater handling of the dependencies such that there remained difficulty with underrepresented classes, such as 'DD', signifying the need for other sources of samples from variably distributed objects.
    \item Effect of Fine-tuning (RQ-3): Fine-tuning LMs on the labeled dataset significantly improved performances of the models, achieving a relatively higher precision-and-recall than that done by LSTMs. However, the noted overfitting on BERT suggests that a larger, diverse dataset would help lower the limit of generalizability. Future work will entail augmentation of the dataset and attempt among many other LMs to sidestep these shortcomings.
\end{enumerate}
Overall, the evaluated LLMs displayed the most promise for smart contract analysis surpassing the traditional ML models in complex vulnerability detection, thus offering promising venues for future investigation. 

\section{Future Work}
This work paves the way for large-scale ML and LLM-based detection of smart contract vulnerabilities, with robustness. These results demonstrate that the method of LLMs can address complexity per smart contracts as DistilBERT and BERT are very specific together thus displaying promising peaks, more than what may be achieved by conventional methods. However, the study is still in progress, and some improvements are expected as \textbf{Hybrid Approach:} Bidirectional Mixing of Labeled \& Unlabeled Data approach \cite{sun2023assbert} in both directions. It contains the dataset expansion which adds more smart contracts as labeled and unlabeled. This is also to minimize the impact of classes that are underrepresented so as to aid in the generalization of our models. \textbf{Exploration of New LLMs:} Evaluating new Large Language Models like GPT-4 \cite{achiam2023gpt}, RoBERTa \cite{liu2019roberta} and additional domain adapted models for their ability to detect or locate vulnerabilities in the context of more complex and contextual code patterns. Semi-Supervised Learning: Working with unlabeled data with semi-supervised learning could help the different LLM models to improve their richer representation and reduce overfitting issues. \textbf{Explainability and Interpretability:} Evaluating the model with Explainable AI \cite{kumar2024digital} will provide a in detailed insight of the model's prediction and classification which will help us to determine how much rely upon the trust on these models. \textbf{Applicability: } In the application-level implementation \cite{onica2023can}, a UX design prototype and backend system architecture for real-time monitoring of vulnerability detection in smart contracts could be developed, enabling practical integration into decentralized applications. Overall, the preliminary results show that LLM can complement traditional ML models in smart contract analysis, providing a good direction for future research and a useful tool to improve blockchain security.

\section{Conclusion}
This research explored the classical machine learning (LSTM) models and the fine-tuned large language models (DistilBERT and BERT) respectively in detecting the vulnerabilities in smart contracts. The outcomes point out that one hand, LSTM models prove to be strong in performance and generalization, and on the other hand, LLMs such as DistilBERT and BERT are better in recognizing complex code patterns and context-specific vulnerabilities. The best overall accuracy was obtained by DistilBERT, while BERT demonstrated some abilities but was slightly overfitted. The results show the benefits of utilizing LLMs for automatic smart contract analysis, moreover, in capturing the dependencies that the typical models may not be able to capture. Even so, the difficulties seen in addressing neglected groups indicate the importance of a more diverse dataset along with enhanced training strategies. The future directions will be data expansion, LLM exploration, and semi-supervised learning for boosting model robustness and generalization. Consequently, this would mean that LLMs that have been opportunely fine-tuned on labeled smart contract data will contribute greatly to vulnerability detection, which in turn opens a path for the development of more reliable and efficient security analysis tools in blockchain systems.


\bibliographystyle{IEEEtran}
\bibliography{bibitem}

\end{document}